\documentclass[aps,prl,twocolumn,superscriptaddress,amsmath]{revtex4-2}

\usepackage{graphicx}
\usepackage{hyperref}
\usepackage{mathrsfs}
\usepackage{bm}
\usepackage{color}
\usepackage{siunitx}
\usepackage[capitalize]{cleveref}
\hypersetup{hypertex=true,
	  colorlinks=true,
	anchorcolor=blue,
	linkcolor=blue,
        citecolor=blue,
	urlcolor=blue}

\begin{document}

\preprint{APS/123-QED}

\title{Spin-to-Charge Conversion Driven by Inverse Chiral-Induced Spin Selectivity}

\author{Tian-Yi Zhang}
\affiliation{International Center for Quantum Materials, School of Physics, Peking University, Beijing 100871, China}

\author{Peng-Yi Liu}
\affiliation{International Center for Quantum Materials, School of Physics, Peking University, Beijing 100871, China}

\author{Ai-Min Guo}
\affiliation{Hunan Key Laboratory for Super-microstructure and Ultrafast Process, School of Physics, Central South University, Changsha 410083, China
}

\author{Qing-Feng Sun}
 \email[]{sunqf@pku.edu.cn}
\affiliation{International Center for Quantum Materials, School of Physics, Peking University, Beijing 100871, China}
\affiliation{Hefei National Laboratory, Hefei 230088, China
}

\begin{abstract}
Chiral molecules have attracted significant multidisciplinary interest and extensive research owing to their remarkable ability to achieve charge-to-spin conversion, known as the chiral-induced spin selectivity (CISS).
A recent experiment has revealed that chiral molecules also exhibit an unexpected capability for spin-to-charge conversion, referred to as the inverse CISS (ICISS), opening unprecedented avenues for the study and application of chiral molecules.
Here, we propose a theoretical model, suggesting that ICISS can be understood in terms of spin-dependent electron deflection induced by the interplay between spin and chiral structure.
Our numerical results are consistent with experimental observations, demonstrating that ICISS persists under strong disorder.
Our model also reproduces the inverse spin Hall effect (ISHE) in this experiment.
Comparative analysis indicates that ICISS exhibits a spin-to-charge conversion behavior that differs from ISHE.
Our work develops a microscopic theoretical model that accounts for the experimentally observed phenomena, and may provide a useful perspective for organic spintronics.
\end{abstract}

\maketitle

\emph{Introduction}\textemdash Spintronics, a vibrant interdisciplinary field bridging physics, chemistry, and biology, has received significant attention through its manipulation of electron spin \cite{Wolf2001}.
A central challenge in spintronics is achieving fast and low-power reversible conversion between charge and spin \cite{Wolf2001,Hirohata2020}.
Conventional approaches utilizing the spin Hall effect (SHE) \cite{Hirsch1999,Sinova2015} and inverse SHE (ISHE) \cite{Saitoh2006,Ando2011} have established a feasible framework where charge current $\bm{J}_c$ and spin current $\bm{J}_s$ become interconvertible.
The SHE generates a transverse spin current from a charge current $\bm{J}_s \propto \bm{J}_c \times \bm{\sigma}_s$, while its reciprocal ISHE produces a transverse charge current from a spin current $\bm{J}_c \propto \bm{J}_s \times \bm{\sigma}_s$, with $\bm{\sigma}_s$ denoting spin polarization direction.
These effects have become powerful tools for both material characterization \cite{Sinova2015} and device implementation \cite{Manipatruni2019}.

Emerging as an attractive complementary route, the chiral-induced spin selectivity (CISS) effect opens a promising avenue for spintronics \cite{Ray1999,Gohler2011,Nakajima2023,Sun2024sciadv,Shiby2025}.
Chirality, a ubiquitous property in nature, has been shown to be intrinsically linked to electron spin \cite{Bloom2024}.
When a spin-unpolarized charge current passes through chiral molecules, it becomes spin-polarized, and the polarization direction is dependent on molecular chirality \cite{Bloom2024,Naaman2019}.
First observed in photoemission experiments with stearoyl lysine \cite{Ray1999}, the CISS has since been robustly confirmed in diverse chiral materials \cite{Gohler2011,Xie2011,Das2025,Bloom2024,Xu2023,Mishra2019,Qian2022,Ko2022,Malatong2023,Goren2025,Al-Bustami2022}.
Meanwhile, numerous theories have been proposed to explain CISS \cite{Guo2012,Guo2014,Alwan2021,Das2022,Zhang2025a,Zhao2025,Liu2025a,Yang2021,Yang2025}, with spin-orbit coupling (SOC) generally acknowledged as playing a key role in the spin-selective process \cite{Guo2012,Guo2014,Alwan2021,Liu2025a,Zhang2025b,Gutierrez2012}.
These experimental and theoretical achievements all demonstrate that CISS holds tremendous research and application prospects \cite{Bloom2024,Xu2023,Ben2017,Aizawa2023,Adhikari2023}.

\begin{figure}[t]
\includegraphics[width=\linewidth]{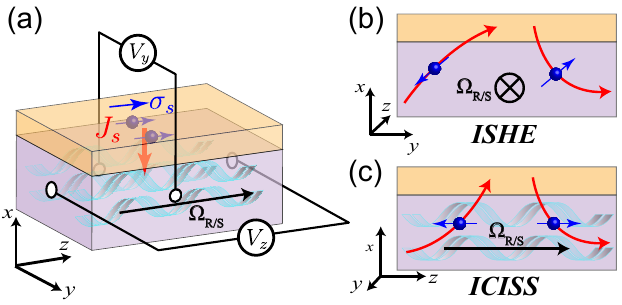}
\caption{\label{fig1}
 (a) Schematic diagram of the experimental configuration.
 The orange and purple boxes represent the spin injector layer and the chiral molecular layer, respectively.
 The injector layer injects a spin current $\bm{J}_s$ (red arrow) with spin polarization direction $\bm{\sigma}_s$ (blue arrow) into the chiral molecular layer.
 The chiral axis $\bm{\Omega}_{R/S}$ is parallel to the $+z$ direction.
 Voltage probes $V_y$ and $V_z$ are connected along the $y$ and $z$ directions of the chiral molecular layer.
 (b-c) Schematic diagram of the ISHE (b) and ICISS (c).
 }
\end{figure}

Following the Onsager reciprocity, the CISS effect possesses a reciprocal effect known as the inverse CISS (ICISS) effect (see End Matter), which is anticipated to enable a novel form of chirality-controlled spin-to-charge conversion \cite{Sun2024}.
While ISHE has been discovered in organic materials \cite{Ando2013,Qiu2015}, and ICISS in inorganic metals \cite{Inui2020,Shiota2021}, experimental observation of ICISS in organic materials has remained elusive until very recently \cite{Sun2024}.
In that study, Sun \textit{et al.} used a solution printing method to fabricate thin films of chiral assembled $\pi$-conjugated polymers \cite{Sun2024,Park2022}.s
They used ferromagnetic resonance to inject a pure spin current $\bm{J}_s$ with the spin polarization direction $\bm{\sigma}_s$ into the polymer film plane.
Two distinct voltage probe measurements were used [see Fig.~\ref{fig1}(a)]: one oriented perpendicular to $\bm{\sigma}_s$ to detect the ISHE signal $V_{ISHE}$, and another parallel to $\bm{\sigma}_s$ to detect the ICISS signal $V_{ICISS}$.
Remarkably, $V_{ISHE}$ was independent of molecular chirality, whereas $V_{ICISS}$ reversed with it.
Let us denote by $\bm{\Omega}_{R/S}$ the geometric chiral axis of the helical polymer, defined by its molecular-scale helix, with subscripts R and S indicating right- and left-handed chirality, respectively.
When $\bm{\sigma}_s$ was oriented in-plane and parallel to $\bm{\Omega}_{R/S}$ \cite{Sun2024}, both $V_{ISHE}$ and $V_{ICISS}$ exhibit substantial values.
When $\bm{\sigma}_s$ was switched from parallel to perpendicular (still in-plane) to $\bm{\Omega}_{R/S}$, a finite $V_{ISHE}$ persisted while $V_{ICISS}$ became small.
This landmark experiment offers an ideal platform to study molecular chirality and spin conversion phenomena \cite{Sun2024}.
However, the observed ICISS requires further evaluation \cite{Nabei2020}, and its theoretical origin remains unclear, underscoring an urgent need for further theoretical elucidation.

In this Letter, we provide a theoretical framework for spin-to-charge conversion in experimentally studied chiral molecular systems.
We demonstrate that both ISHE and ICISS originate from SOC within the system, with ICISS arising from the combined effect of the chiral structure and SOC.
By constructing a theoretical model, we numerically reproduce the main experimental features including ISHE and ICISS, and show their persistence even under strong disorder.
We also show a fundamental distinction between ISHE and ICISS: while ISHE satisfies $\bm{J}_c \propto \bm{J}_s \times \bm{\sigma}_s$, ICISS exhibits $\bm{J}_c \propto \bm{\Omega}_{R/S}$.
Our work provides a microscopic interpretation of ICISS and suggests that chiral molecules can serve as a promising platform for spin conversion.

\emph{Theoretical model}\textemdash We first construct a theoretical model corresponding to the experimental configuration \cite{Sun2024} (see ``Model Hamiltonian'' section in End Matter).
As shown in Fig.~\ref{fig1}(a) and Fig. S1 in Sec. S1 in Supplemental Material \cite{SM}, the orange box represents the spin injector layer.
Directly beneath it, the purple box denotes the chiral molecular layer, which consists of an assembly of polymers possessing chiral structures \cite{Sun2024,Park2022,Song2007,Frederiksen2009}.
The chiral molecular layer has size parameters $L_x=5$ ($x$ direction thickness) and $L_y=25$ ($y$ direction width), meaning there are $L_x \times L_y$ chiral molecules.
We use our previously developed model \cite{Guo2012,Guo2014}, which includes SOC and chiral structures, to simulate the single chiral molecule.
This model successfully accounts for various CISS-related phenomena \cite{Guo2014,Zhang2025a,Liu2025a,Zhang2025b,Liu2025b}, demonstrating its robust explanatory power.
Each molecule has a length $L_z=25$ along $\bm{\Omega}_{R/S}$.
Voltage probes are attached along the $y$ and $z$ directions of the chiral molecular layer, with their voltage differences denoted $V_y$ and $V_z$, respectively.
Here we adopt a coordinate system where the $(x, y, z)$ axes correspond respectively to the $(z, x, y)$ axes in the experiment \cite{Sun2024}.
This labeling preserves formal consistency with prior theoretical works \cite{Guo2012,Guo2014} and does not affect physical conclusions, provided the measurement geometry matches that of the experiment.

Next, we simulate the experimental spin-pumping process induced by ferromagnetic resonance.
We model the experimental spin injection \cite{Moharana2025} by assigning the chemical potentials $\pm eV_0$ to the two spin species in the injector layer \cite{Liu2025b,Brataas2002,Wang2004}.
In the experiment, the microwave excitation is at $f = 6 \ \mathrm{GHz}$, corresponding to $V_0 \approx hf/2e=12.4 \ \mathrm{\mu V}$ \cite{Liu2025b}.
Under this spin bias, a pure spin current $\bm{J}_s$ with polarization direction $\bm{\sigma}_s$ is injected into the chiral molecular layer [Fig.~\ref{fig1}(a)] \cite{Sun2005a}.

We then analyze the physical origins of the ISHE and ICISS.
Both effects are induced by the SOC within the system.
One contribution to the SOC originates from the interfacial electric field $\bm{E}_x$ (along the $x$-axis) at the heterojunction formed between two layers \cite{Edelstein1990}.
This SOC takes the form $H_{so}^E=e\hbar/(4m_e^2 c^2)[\bm{E}_x \cdot(\bm{\sigma}\times \bm{p})]$ \cite{Bychkov1984}, where $m_e$, $c$, $\bm{\sigma}=(\sigma_x,\sigma_y,\sigma_z)$ and $\bm{p}$ represent electron mass, speed of light, Pauli matrices, and electron momentum, respectively.
Another contribution to the SOC is the intra-molecular SOC \cite{Guo2012,Guo2014,Pan2016,Wang2024,Steele2013,Sun2005b}, which takes the form $\hbar/(4m_e^2 c^2)[\nabla V \cdot (\bm{\sigma}\times \bm{p})]$, where $V$ is the electrostatic potential of helically stacked monomers.
These SOC mechanisms induce spin-dependent electron deflection, generating ISHE and ICISS.

Figures \ref{fig1}(b) and \ref{fig1}(c) schematically depict electron transport behaviors of ISHE and ICISS, respectively.
In both ISHE and ICISS, what is injected from the injector layer is a spin current--that is, spin-up electrons are injected while spin-down electrons return.
In ISHE [Fig.~\ref{fig1}(b)], electrons with $+z$ ($-z$) aligned spins are deflected toward the $+y$ surface (return from the $-y$ surface) due to SOC, as illustrated by the red curved arrows.
This creates a transverse charge accumulation that manifests as a non-zero ISHE signal $V_y$.
In ICISS [Fig.~\ref{fig1}(c)], the spin-up and spin-down transmission probabilities for electrons from the injector layer to the $z$-oriented surfaces are different because of the chiral structure and SOC \cite{Guo2012,Pan2016,Sun2005b,Chen2024}, and the transmission difference depends on $\bm{\sigma}_s$ and $\bm{\Omega}_{R/S}$ \cite{Guo2012,Guo2014,Zhang2025a,Pan2016} (see Sec. S2 for details \cite{SM}).
As illustrated in Fig.~\ref{fig1}(c), electrons with $+z$ ($-z$) aligned spins  preferentially propagate toward the $+z$ surface (return from the $-z$ surface).
This leads to a charge transfer along $\bm{\Omega}_{R/S}$, giving rise to ICISS.

\begin{figure}[t]
\includegraphics[width=\linewidth]{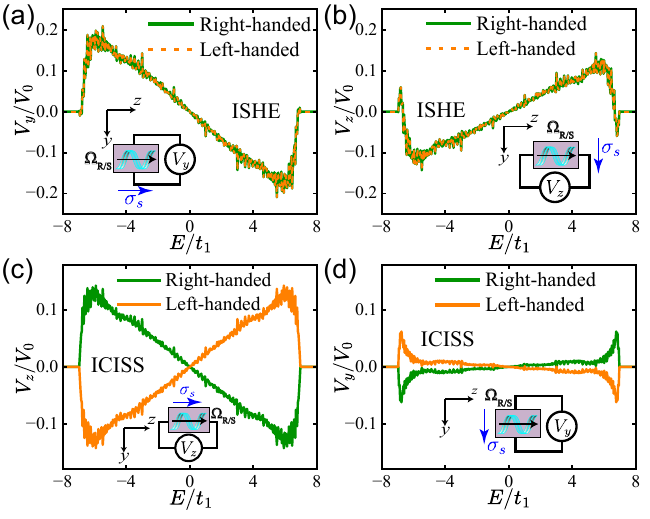}
\caption{\label{fig2}
 (a-b) ISHE voltage of the right-/left-handed (green/orange dashed line) molecules versus Fermi energy $E$.
 (c-d) ICISS voltage of the right-/left-handed (green/orange line) molecules versus $E$.
 $V_y$ ($V_z$) is measured along the $+y$ ($+z$) direction.
 $\bm{\sigma}_s$ is along the $+z$ (a,c) and $+y$ (b,d) direction.
 The inset of each panel illustrates the corresponding measurement configuration, where $\bm{\sigma}_s$ is marked by the blue arrow, $\bm{\Omega}_{R/S}$ is along the $+z$ direction.
 Parameters used here are provided in ``Model parameters'' section in End Matter.}
\end{figure}

\emph{Transport behavior of ISHE and ICISS}\textemdash Next, we sequentially investigate the transport behavior of each experimental configuration using the non-equilibrium Green's function method \cite{Xing2007,Lee1981,Datta1995} (see ``Transport formulation'' section in End Matter).
In all configurations, $\bm{\Omega}_{R/S}$ is aligned along the $z$ direction, and the spin current $\bm{J}_s$ flows in the $-x$ direction.

In the first configuration shown in the inset of Fig.~\ref{fig2}(a), $\bm{\sigma}_s$ is along the $+z$ direction, and $V_y$ is measured along the $y$ direction.
Since the voltage measurement direction is perpendicular to $\bm{\sigma}_s$, this corresponds to the ISHE voltage \cite{Sun2024}.
The green line and orange dashed line in Fig.~\ref{fig2}(a) show $V_y$ versus the system's Fermi energy $E$ for right-handed and left-handed molecules, respectively.
Several important features can be identified:
(i) A non-zero $V_y$ appears when $E$ lies within the energy band.
$|V_y|$ increases near the band edges, and $V_y=0$ outside the band.
(ii) $V_y$ also exhibits oscillating behavior, with the oscillations intensifying near the band edges $|E|=6t_1$ ($t_1$ is the intra-chain hopping integral \cite{Yan2002}, see End Matter).
These oscillations are caused by quantum interference \cite{Pan2016}.
(iii) $V_y>0$ for electron-like carriers ($-6t_1<E<0$) and $V_y<0$ for hole-like carriers ($0<E<6t_1$).
(iv) The $V_y$ curve possesses central symmetry with respect to the point $(0,0)$, i.e., $V_y (-E)=-V_y (E)$, which originates from electron-hole symmetry \cite{Pan2016}.
(v) Crucially, when the molecular chirality is reversed (changing structure parameter as $\theta_{mol} \to \pi-\theta_{mol}$, $\Delta \varphi \to -\Delta \varphi$, see End Matter), the two curves completely coincide.
This demonstrates that ISHE is entirely independent of molecular chirality (see details in Sec. S3 \cite{SM}).
Here, $V_y$ ranges from approximately up to $0.2V_0$, i.e., a few $\mathrm{\mu V}$, which matches the magnitude of the voltages measured experimentally \cite{Sun2024}.

The inset of Fig.~\ref{fig2}(b) shows the second experimental configuration, where $\bm{\sigma}_s$ is along the $+y$ direction, and the voltage $V_z$ is measured along the $z$ direction.
Since the measurement direction remains perpendicular to $\bm{\sigma}_s$, this also corresponds to the ISHE voltage \cite{Sun2024}.
Fig.~\ref{fig2}(b) shows $V_z$ versus $E$.
It reveals that the $z$ direction also produces an ISHE signal similar to Fig.~\ref{fig2}(a), except for the difference in feature (iii):
Here, $V_z<0$ for electron-like carriers ($-6t_1<E<0$) and $V_z>0$ for hole-like carriers ($0<E<6t_1$).
According to $\bm{J}_c \propto \bm{J}_s \times \bm{\sigma}_s$, electron-like carriers with spin polarized along $+y$ are deflected towards the $-z$ direction, which is opposite to the $V_z$ measurement direction ($+z$), so $V_z<0$.
It is shown in Figs.~\ref{fig2} (a,b) that the ISHE signal appears as long as the measurement direction is perpendicular to $\bm{\sigma}_s$, regardless of whether $\bm{\sigma}_s$ is parallel or perpendicular to $\bm{\Omega}_{R/S}$.

Next, we study the ICISS according to the experimental configurations \cite{Sun2024}.
The inset in Fig.~\ref{fig2}(c) shows the experimental setup for measuring the $z$ direction ICISS.
There, both $\bm{\sigma}_s$ and the voltage measurement direction are along the $+z$ direction, parallel to $\bm{\Omega}_{R/S}$.
Fig.~\ref{fig2}(c) shows $V_z$ versus $E$.
We observe that a non-zero ICISS signal shows up, exhibiting both similarities to and distinctions from ISHE.
First, ICISS exhibits features (i, ii, iv) of ISHE shown in Fig.~\ref{fig2}(a).
However, in ICISS, the electron deflection occurs along $\bm{\Omega}_{R/S}$, so that the resultant voltage is in the $z$ direction, which differs from ISHE.
Moreover, when the molecular chirality is reversed, the ICISS signal $V_z$ also reverses sign to $-V_z$.
This demonstrates that ICISS is highly dependent on molecular chirality \cite{SM}, consistent with experimental results \cite{Sun2024}.
Like CISS, ICISS also exists at the single-molecule level.
In Fig. S2 \cite{SM}, we show the ICISS voltage generated across a single molecule ($L_x=L_y=1, \ L_z=25$). The magnitude in Fig. S2 is similar to that in Fig.~\ref{fig2}(c), consistent with the expected voltage per molecule in a parallel configuration.

The inset in Fig.~\ref{fig2}(d) shows the experimental setup for measuring the $y$ direction ICISS \cite{Sun2024}.
Here, both the voltage measurement direction and $\bm{\sigma}_s$ are perpendicular to $\bm{\Omega}_{R/S}$.
Fig.~\ref{fig2}(d) reveals that $V_y$ is small, demonstrating that ICISS, like CISS, predominantly occurs along $\bm{\Omega}_{R/S}$.
While the ICISS signal is small, it is non-zero and also reverses its sign when the molecular chirality is reversed.
This was indeed observed experimentally \cite{Sun2024}, indicating agreement between our model and the experimental observations.

In real experimental systems, disorder inevitably arises owing to misalignment, defects, or impurities.
In Sec. S4 \cite{SM}, we analyze the effects of disorder on the four voltage configurations shown in Figs.~\ref{fig2}(a-d).
Beyond disorder robustness, in Sec. S5 we also verify the reciprocal relationship between CISS and ICISS by calculating their Onsager coefficients \cite{SM}.
Moreover, the out-of-plane ICISS effect reported recently \cite{Dong2025} is also investigated in Sec. S6 \cite{SM},
and our theory provides a unified explanation for both in-plane and out-of-plane ICISS.
In addition, the CISS effect is mainly observed in low-conductivity systems, whereas the ICISS effect has also been reported in more conductive systems. These are discussed in Sec. S6 \cite{SM}.

\begin{figure}[t]
\includegraphics[width=\linewidth]{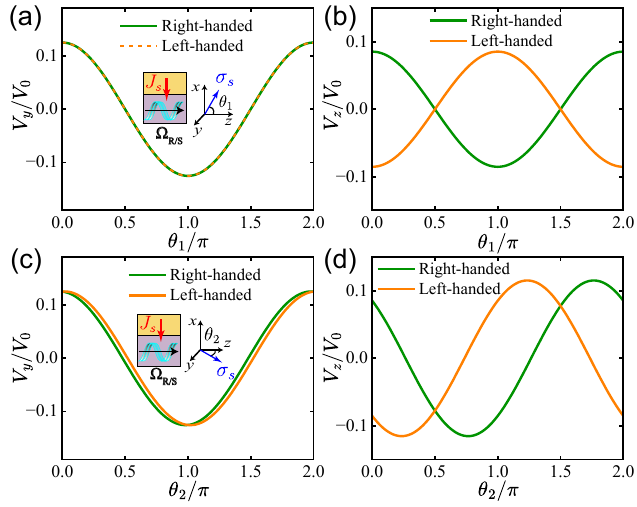}
\caption{\label{fig3}
 (a-b) Voltage variation along the $y$ (a) and $z$ (b) directions as $\bm{\sigma}_s$ rotates within the $xz$-plane (angle $\theta_1$ relative to the $z$-axis, shown in the inset of a).
 $\bm{J}_s$ is the direction of spin current flow.
 (c-d) Voltage variation along the $y$ (c) and $z$ (d) directions as $\bm{\sigma}_s$ rotates within the $yz$-plane (angle $\theta_2$ relative to the $z$-axis, shown in the inset of c).
 $\bm{\Omega}_{R/S}$ is along the $z$ direction, $E=-4t_1$, and other parameters used here are provided in ``Model parameters'' section in End Matter.}
\end{figure}

\emph{Spin-polarization-angle dependence of ISHE and ICISS}\textemdash We further investigate the dependence of ISHE and ICISS on $\bm{\sigma}_s$ of the injected spin current, following the experimental configurations \cite{Sun2024}.
$\bm{\Omega}_{R/S}$ is always aligned along the $z$ direction, and energy is fixed at $E=-4t_1$.
We first investigate the case where $\bm{\sigma}_s$ lies within the $xz$ plane, shown in the inset of Fig.~\ref{fig3}(a).
Here, the angle between $\bm{\sigma}_s$ and $+z$ axis is denoted as $\theta_1$.
Fig.~\ref{fig3}(a) shows $V_y$ versus $\theta_1$.
Since the $yz$ plane component of the spin lies along the $z$ direction, and the voltage is measured along the $y$ direction, only ISHE is present here.
Because the component of $\bm{\sigma}_s$ within the $yz$ plane varies as $\cos \theta_1$, $V_y$ follows $V_y(\theta_1)=V_y (0) \cos \theta_1$.
Crucially, the $V_y$ curves for molecules of opposite chirality completely coincide.
Fig.~\ref{fig3}(b) shows $V_z$ versus $\theta_1$.
Since both the measurement direction and the projection direction of the spin within the $yz$ plane are along $z$, only ICISS is present here.
The projection component of the $\bm{\sigma}_s$ along $\bm{\Omega}_{R/S}$ also varies as $\cos \theta_1$.
Therefore, the $V_z$ follows $V_z(\theta_1)=V_z(0)\cos \theta_1$.
Significantly, the $V_z$ curves for molecules of opposite chirality have opposite signs.
In experiments, the injected spin polarization may not be aligned with the external magnetic field.
This is discussed in Sec. S7 \cite{SM}.

All preceding results [Figs.~\ref{fig2}, \ref{fig3}(a-b)] derive from theoretical calculations replicating experimental measurement configurations, showing agreement with experiment \cite{Sun2024}.
We now extend our investigation to the experimentally uncharted regime where $\bm{\sigma}_s$ rotates within the $yz$-plane.
In Figs.~\ref{fig3}(c) and (d), the angle between $\bm{\sigma}_s$ and the $+z$ direction is denoted as $\theta_2$, shown in the inset of Fig.~\ref{fig3}(c).
Fig.~\ref{fig3}(c) shows $V_y$ versus $\theta_2$.
It shows a small phase shift between the two curves, because both ISHE and ICISS contribute to the voltage.
When the molecular chirality changes, the contribution from ICISS reverses sign, while that from ISHE remains constant, resulting in the difference between the two curves.
However, because the voltage is measured perpendicular to $\bm{\Omega}_{R/S}$, the contribution of ICISS is small, so the shift is small.
Fig.~\ref{fig3}(d) shows $V_z$ versus $\theta_2$.
It shows a large phase shift between the two curves.
In this case, because the voltage is measured parallel to $\bm{\Omega}_{R/S}$, the contribution of ICISS is large, so the shift is large.
The results in Fig.~\ref{fig3} collectively demonstrate that within the chiral molecular layer, ISHE and ICISS can be studied both individually [Figs.~\ref{fig3}(a,b)] and simultaneously [Figs.~\ref{fig3}(c,d)].

\begin{figure}[t]
\includegraphics[width=\linewidth]{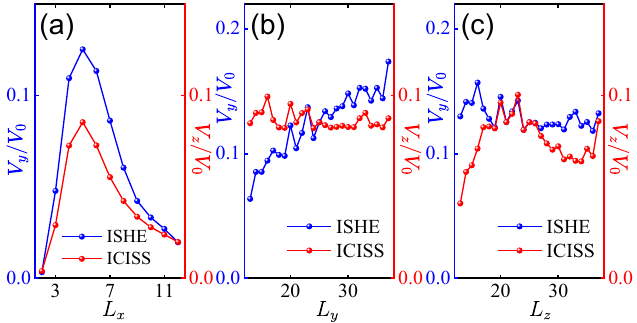}
\caption{\label{fig4}
 (a-c) Dependence of ISHE and ICISS, $V_y$ and $V_z$, on molecular layer thickness $L_x$ in (a), on molecular layer width $L_y$ in (b), and on molecular length $L_z$ in (c).
 The molecular dimensions are fixed at $L_x=5$ and  $L_y=L_z=25$, unless they serve as the horizontal axis (variable).
 $E=-4t_1$, the molecule is right-handed, with $\bm{\Omega}_{R/S}$ along the $z$ direction.
 $\bm{\sigma}_s$ is oriented along the $+z$ direction.
 Hence, $V_y$ and $V_z$ correspond to the ISHE and ICISS voltages, respectively.
 }
\end{figure}

\emph{Size dependence of ISHE and ICISS}\textemdash Finally, we investigate the size dependence of ISHE and ICISS in the molecular layer with respect to the size parameters $L_x$, $L_y$, and $L_z$.
In Fig.~\ref{fig4}, $\bm{\Omega}_{R/S}$ and $\bm{\sigma}_s$ are along the $z$ direction, so $V_y$ and $V_z$ represent ISHE and ICISS voltages, respectively.
Fig.~\ref{fig4}(a) shows the variation of $V_y$ and $V_z$ with thickness $L_x$, revealing that both $V_y$ and $V_z$ initially increase with $L_x$, reach a maximum, and decrease inversely with further increases in $L_x$.
Fig.~\ref{fig4}(b) shows the variation of $V_y$ and $V_z$ with width $L_y$, revealing that $V_y$ increases with $L_y$, and $V_z$ remains almost constant regardless of $L_y$; both signals exhibit minor fluctuation behavior.
The increase of $V_y$ with $L_y$ matches ISHE experimental observations \cite{Ando2011}, and our results therefore predict that ICISS is independent of the layer width.
Fig.~\ref{fig4}(c) shows the variation of $V_y$ and $V_z$ with molecular length $L_z$, revealing that $V_y$ remains almost constant for larger $L_z$, while $V_z$ increases with $L_z$ before saturating; both signals exhibit minor fluctuation behavior.
The observed length dependence of ISHE well matches experimental findings \cite{Ando2011}, and the length dependence of ICISS closely resembles the variation of spin polarization with molecular length in CISS \cite{Guo2014,Gutierrez2012,Mishra2020}.
Therefore, our theory predicts that ICISS shares the same molecular length dependence with CISS.

\emph{Conclusions}\textemdash In conclusion, we propose a theoretical model based on the experimental configuration \cite{Sun2024}, and obtain ISHE and ICISS consistent with experimental results.
The experimental observations originate from the SOC arising from the heterojunction and the helically stacked monomers.
While ISHE satisfies $\bm{J}_c \propto \bm{J}_s \times \bm{\sigma}_s$, ICISS satisfies $\bm{J}_c \propto \bm{\Omega}_{R/S}$.
Both effects persist under strong disorder.
We also predict that ISHE and ICISS can contribute simultaneously upon variation of $\bm{\sigma}_s$ within the chiral molecular plane.
Finally, ISHE and ICISS show distinct dependencies on the system size.
We provide a microscopic interpretation of the ICISS effect and may offer guidance for achieving more efficient applications of ICISS in practice (details are provided in Sec. S8 \cite{SM}).

\begin{acknowledgments}
\emph{Acknowledgments}\textemdash This work was financially supported by the National Key R and D Program of China (Grant No. 2024YFA1409002),
the National Natural Science Foundation of China (Grant No. 12374034 and No. 12274466),
Quantum Science and Technology-National Science and Technology Major Project (2021ZD0302403),
and the Hunan Provincial Science Fund for Distinguished Young Scholars (2023JJ10058).
We also acknowledge the High-performance Computing Platform of Peking University for providing computational resources.
\end{acknowledgments}



\begin{onecolumngrid}
\subsection{End Matter}
\end{onecolumngrid}

\setcounter{equation}{0}
\renewcommand{\theequation}{A\arabic{equation}}

\twocolumngrid

\emph{Deriving ICISS from Onsager reciprocity} \textemdash The relationship between CISS and ICISS can be formally derived from the Onsager reciprocity.
Within the linear response regime of transport theory, the charge current ($I_c$), spin current ($I_s$), charge voltage ($V_c$), and spin voltage ($V_s$) are related by the constitutive relation: $I_{i,\mu}=\sum_j G_{ij}^{\mu \nu} V_{j,\nu}$, where $\mu, \nu$ denote the charge ($c$) or spin ($s$) indices, $i,j$ are electrode indices, and $G_{ij}^{\mu \nu}$ is the corresponding conductance coefficient.
According to the Onsager reciprocity relation, under time-reversal symmetry \cite{Xing2007}, the relation $G_{ij}^{sc}=-G_{ji}^{cs}$ holds.
In the CISS effect, a charge voltage can generate a spin current, implying a non-zero $G_{ij}^{sc}$.
The Onsager relation then guarantees that a spin current can drive a charge current in the same system (i.e., non-zero $G_{ji}^{cs}$) with the same magnitude, which is the ICISS effect.
Furthermore, since the conductance coefficients for right-handed ($R$) and left-handed ($S$) systems in CISS satisfy $G_{ij}^{sc} (R)=-G_{ij}^{sc} (S)$, the Onsager relation immediately implies $G_{ij}^{cs} (R)=-G_{ij}^{cs} (S)$ in ICISS.
These formally establish the relation between CISS and ICISS.

\emph{Model Hamiltonian} \textemdash Based on the experimental configuration, we construct our Hamiltonian model.
When probing voltage $V_y$ along $y$ direction, the Hamiltonian is $H_{V_y}=H_{mol-layer}+H_{in}+H_y^{vol}$, and when probing voltage $V_z$ along $z$ direction, the Hamiltonian is $H_{V_z}=H_{mol-layer}+H_{in}+H_z^{vol}$.
Here, $H_{mol-layer}$ describes the chiral molecular layer, and $H_{in}$ describes the spin injector layer and its coupling to the chiral molecular layer.
$H_y^{vol}$ and $H_z^{vol}$ describe the electrodes and their coupling along $y$ and $z$ directions, respectively.
When calculating the ICISS for a single molecule (i.e., setting $L_x=L_y=1$), we additionally couple a normal electrode $H_{sub}$ beneath the molecule and attach virtual electrodes $H_d$ to the molecule to enable proper spin current injection.
Next, we detail the Hamiltonian for the chiral molecular layer, $H_{mol-layer}$,
while the Hamiltonians for other parts can be found in the Supplemental Material \cite{SM}.

The Hamiltonian of chiral molecular layer is $H_{mol-layer}=H_z^{mol}+H_{xy}^{mol}+H_{yz}^E$, where $H_z^{mol}$, $H_{xy}^{mol}$ and $H_{yz}^E$ describe the single nanofiber, electron hopping along the $x$ and $y$ directions, and SOC originating from the heterojunction, respectively.
The chiral molecular layer is formed by stacking multiple nanofibers along the $x$ and $y$ directions.
Each nanofiber consists of two intertwined helical single chains oriented along the $z$-axis.
The Hamiltonian for a single nanofiber is \cite{Guo2012,Guo2014}:
\begin{equation}
	\begin{aligned}
	H_z^{mol}&=\sum_{n_x=1}^{L_x} \sum_{n_y=1}^{L_y} \sum_{n_z=1}^{L_z-1} \sum_{j=1}^2 \big[t_1 c_{j,n_x,n_y,n_z}^{\dagger} c_{j,n_x,n_y,n_z+1}\\
	&+i t_{so} c_{j,n_x,n_y,n_z}^{\dagger} (\sigma_{n_z}^{mol}+\sigma_{n_z+1}^{mol} ) c_{j,n_x,n_y,n_z+1}+h.c. \big]\\
	&+\sum_{n_x=1}^{L_x} \sum_{n_y=1}^{L_y} \sum_{n_z=1}^{L_z} (t_2 c_{1,n_x,n_y,n_z}^{\dagger} c_{2,n_x,n_y,n_z}+h.c.),
	\end{aligned}
\end{equation}
where $c_{j,n_x,n_y,n_z}^\dagger=(c_{j,n_x,n_y,n_z,\uparrow}^\dagger,c_{j,n_x,n_y,n_z,\downarrow}^\dagger )$ is the creation operator at site $(n_x,n_y,n_z)$ of the $j$th chain, with $n_x$, $n_y$ and $n_z$ labeling the monomers in the $x$, $y$ and $z$ direction.
This layer has thickness $L_x$, width $L_y$, and each molecule has a length $L_z$.
$t_1$ is the intra-chain hopping integral, $t_2$ is the inter-chain hopping integral, and $t_{so}$ is the intrinsic molecular SOC.
$\sigma_{n_z+1}^{mol}=-\sigma_x  \sin(n_z \Delta \varphi)  \sin \theta_{mol} +\sigma_y  \cos(n_z \Delta \varphi)  \sin \theta_{mol} + \sigma_z  \cos \theta_{mol} $, where $\sigma_{x,y,z}$ are the Pauli matrices.
$\theta_{mol}=\arctan[h/(2 \pi R)]$ is the helix angle of molecule with helical pitch $h$ and helical radius $R$, and $\Delta \varphi$ is the twist angle between neighboring sites.
The first term describes the ``through-bond'' motion of electrons along each helical single chain \cite{Song2007,Frederiksen2009}.
This Hamiltonian is obtained through the same discretization procedure as in our previous works \cite{Guo2012,Guo2014}, and maintaining formal consistency.
The second term describes the ``chain-to-chain'' hopping between the two intertwined chains (see Fig. S1 \cite{SM}).
Due to the close proximity of the chains, only the nearest-neighbor hopping $t_2$ between adjacent sites is considered.

The Hamiltonian of inter-nanofiber hopping of electrons is:
\begin{equation}
	\begin{aligned}
	H_{xy}^{mol}&=\sum_{n_x=1}^{L_x-1} \sum_{n_y=1}^{L_y} \sum_{n_z=1}^{L_z} \sum_{j=1}^2 t_2 c_{j,n_x,n_y,n_z}^\dagger c_{j,n_x+1,n_y,n_z} \\
	&+\sum_{n_x=1}^{L_x} \sum_{n_y=1}^{L_y-1} \sum_{n_z=1}^{L_z} \sum_{j=1}^2 t_2 c_{j,n_x,n_y,n_z}^\dagger c_{j,n_x,n_y+1,n_z}+h.c.
	\end{aligned}
\end{equation}
The first term describes the inter-nanofiber hopping of electrons in the $x$ direction, while the second term describes the inter-nanofiber hopping in the $y$ direction.
In the experiment, the distance between adjacent helical nanofiber is small, and the magnitude of inter-nanofiber hopping decreases exponentially rapidly with distance.
Therefore, only the nearest-neighbor inter-nanofiber hopping perpendicular to the helical axis direction is considered here.

The SOC induced by the electric field parallel to the $x$ axis is described by:
\begin{equation}
	\begin{aligned}
	&H_{yz}^E =\\
	&\sum_{n_x=1}^{L_x} \sum_{n_y=1}^{L_y} \sum_{n_z=1}^{L_z-1} \sum_{j=1}^2 it_{so}^E c_{j,n_x,n_y,n_z}^\dagger (\sigma_{n_z}^E+\sigma_{n_z+1}^E ) c_{j,n_x,n_y,n_z+1}\\
	&-\sum_{n_x=1}^{L_x} \sum_{n_y=1}^{L_y} \sum_{n_z=1}^{L_z} 2i t_{so}^E c_{1,n_x,n_y,n_z}^\dagger \sigma_y c_{2,n_x,n_y,n_z}\\
	&+\sum_{n_x=1}^{L_x} \sum_{n_y=1}^{L_y-1} \sum_{n_z=1}^{L_z} \sum_{j=1}^2 2it_{so}^E c_{j,n_x,n_y,n_z}^\dagger \sigma_z c_{j,n_x,n_y+1,n_z}+h.c.,
	\end{aligned}
\end{equation}
where $t_{so}^E$ represents the corresponding SOC strength, and $\sigma_{n_z+1}^E=-\sigma_y \sin \theta_{mol}-\sigma_z \cos(n_z \Delta \varphi) \cos \theta_{mol}$.
This Hamiltonian is obtained by discretizing $H_{so}^E=e\hbar/(4m_e^2 c^2)[\bm{E}_x \cdot (\bm{\sigma}\times \bm{p})]$, following the same procedure applied to the chiral molecules.
The first term describes the SOC experienced by electrons moving within a single chain due to the heterojunction's electric field.
The second term describes the SOC experienced by electrons moving between the two chains within the same nanofiber.
The third term describes the SOC experienced by electrons during inter-nanofiber movement along the $y$ direction.

The Hamiltonian of the chiral molecular layer $H_{mol-layer}$ preserves the electron-hole symmetry \cite{Pan2016}, i.e. $H_{mol-layer}$ remains invariant under the transformation $c_{j,n_x,n_y,n_z,\uparrow}\to (-1)^{j+n_x+n_y+n_z} c_{j,n_x,n_y,n_z,\downarrow}^\dagger$, $c_{j,n_x,n_y,n_z,\downarrow}\to (-1)^{j+n_x+n_y+n_z+1} c_{j,n_x,n_y,n_z,\uparrow}^\dagger$.
This property leads to $V_y (-E)=-V_y (E)$, and $V_z (-E)=-V_z (E)$ in Fig.~\ref{fig2}. When disorder is taken into account, the Hamiltonian needs to add an additional term: $H_{dis}=\sum_{n_x=1}^{L_x} \sum_{n_y=1}^{L_y} \sum_{n_z=1}^{L_z} \sum_{j=1}^2 w_{j,n_x,n_y,n_z} c_{j,n_x,n_y,n_z}^\dagger c_{j,n_x,n_y,n_z}$, where the on-site disorder energy $w_{j,n_x,n_y,n_z}$ is uniformly distributed in the range $[-W/2,W/2]$ with the disorder strength $W$.

\emph{Transport formulation} \textemdash The voltage measurement configurations along the $y$ and $z$ directions are calculated similarly using the non-equilibrium Green's function method. In the following we label the electrodes with $p$, following convention:
(1) For the $y$ direction voltage measurement, the Hamiltonian is $H_{V_y}=H_{mol-layer}+H_{in}+H_y^{vol}$. The labels are $p=in,\pm y$, corresponding to the electrodes described by $H_{in}$ and $H_y^{mol}$, respectively.
(2) For the $z$ direction voltage measurement, the Hamiltonian is $H_{V_z}=H_{mol-layer}+H_{in}+H_z^{vol}$. The labels are $p=in,\pm z$, corresponding to the electrodes described by $H_{in}$ and $H_z^{vol}$, respectively.
(3) For the single molecule ICISS measurement, the Hamiltonian is $H_{single}=H_{mol-layer}+H_{in}+H_z^{vol}+H_{sub}+H_d$. The labels are $p=in,\pm z,sub,d,$ corresponding to the electrodes described by $H_{in}$, $H_z^{vol}$, $H_{sub}$, and $H_d$, respectively.
In this measurement, we set $L_x=L_y=1$.

According to the Landauer-Büttiker formula \cite{Datta1995}, the current with spin $s=\uparrow, \downarrow$ from electrode $p$ to the chiral molecular layer is given by:
\begin{equation}
	\begin{aligned}
	I_{p,s}=\frac{e^2}{h} \sum_{qs'} T_{ps,qs'} (V_{p,s}-V_{q,s'}),
	\end{aligned}\label{LB_formula}
\end{equation}
where $T_{ps,qs'}=\text{Tr}[\mathbf{\Gamma}_{ps'} \mathbf{G}^r \mathbf{\Gamma}_{qs'} \mathbf{G}^a]$ is the transmission coefficient from electrode $q$ with spin $s'$ to the electrode $p$ with spin $s$.
The summation in this formula includes the case where $q\neq p$ and $q=p$ (but $s'=-s$), and the term $T_{ps,p-s}$ describes the spin-flip reflection at the same electrode $p$.
The retarded Green's function $\mathbf{G}^r=[\mathbf{G}^a]^\dagger=[E\mathbf{I}-\mathbf{H}_{mol-layer}-\sum_{ps} \mathbf{\Sigma}_{ps}^r]^{-1}$, the linewidth function $\mathbf{\Gamma}_{ps}=i[\mathbf{\Sigma}_{ps}^r-\mathbf{\Sigma}_{ps}^a]$, and $\mathbf{I}$ is the identity matrix.
The bold letters represent the matrix under the tight-binding representation.
$E$ is the Fermi energy, $\mathbf{\Sigma}_{ps}^r=[\mathbf{\Sigma}_{ps}^a]^\dagger$ is the retarded self-energy due to the coupling to the electrode $p$ with spin $s$.
For the spin injector layer, the self-energy of spin-$s$ (in arbitrary direction) can be calculated numerically \cite{Xing2007,Lee1981}.
For the electrodes along the $y$ and $z$ directions, $\mathbf{\Sigma}_{ps}^r=-i\Gamma_p/2$ is energy independent, with $\Gamma_p$ the linewidth parameter.

Considering the chemical potential splitting of different spins in the spin injector layer, the spin voltage is obtained as $V_{in,s}=-V_{in,-s}=V_0$.
For the voltage probes in the $y$ and $z$ directions, $V_{p,s}=V_{p,-s}$.
Since there is no net charge current for these voltage probes, we have $I_{p,s}+I_{p,-s}=0$.
Under these boundary conditions, the current and voltage at each electrode can be derived using Eq. (\ref{LB_formula}), and the measured voltage is given by $V_y=V_{+y}-V_{-y}$ and $V_z=V_{+z}-V_{-z}$.

\emph{Model parameters} \textemdash Within the chiral molecular layer, experimental measurements show identical carrier mobility parallel and perpendicular to the chiral axis $\bm{\Omega}_{R/S}$ \cite{Sun2024}.
This indicates equivalent intra-nanofiber and inter-nanofiber hopping energies \cite{Datta1995}.
Therefore, we set $t_1=t_2$ and take $t_1$ as the energy unit.
The structural parameters of single helical chain adapted from the experiment \cite{Sun2024,Park2022} are the radius $R = 10 \ \text{nm}$, and the helical pitch $h = 40 \ \text{nm}$.
The twist angle is $\Delta \varphi= \pi/12$.
Limited by computational resources (the entire system is three-dimensional), we simulate a system of relatively small size: $L_x=5$, $L_y=25$, $L_z=25$.
The SOC strengths \cite{Sun2005b} are: $t_{so}=0.03 t_1$, $t_{so}^E=0.05t_1$.
In organic molecular systems, the hopping integral $t_1$ is typically on the order of 100 meV \cite{Guo2012,Yan2002}, which places $t_{so}$ of 3 meV well within the experimentally reasonable range for such systems \cite{Steele2013,Evers2022,Jhang2010}.
For the voltage probes along the $y$ and $z$ directions, the linewidth parameters are fixed as $\Gamma_{+y}=\Gamma_{-y}=\Gamma_{+z}=\Gamma_{-z}=t_1$.
For the virtual electrodes, $\Gamma_d=0.1t_1$.
The strength of disorder $W=t_1$ in Figs. S4(c-f) \cite{SM}.
The above-mentioned values of parameters are used throughout this paper except for specific indications in the figure.


\begin{thebibliography}{100}

	\bibitem{Wolf2001}
	S. A. Wolf, D. D. Awschalom, R. A. Buhrman, J. M. Daughton, S. Von Molnár, M. L. Roukes, A. Y. Chtchelkanova, and D. M. Treger, Spintronics: A spin-based electronics vision for the future, Science \textbf{294}, 1488 (2001).

	\bibitem{Hirohata2020}
	A. Hirohata, K. Yamada, Y. Nakatani, I.-L. Prejbeanu, B. Diény, P. Pirro, and B. Hillebrands, Review on spintronics: Principles and device applications, J. Magn. Magn. Mater. \textbf{509}, 166711 (2020).

	\bibitem{Hirsch1999}
	J. E. Hirsch, Spin Hall effect, Phys. Rev. Lett. \textbf{83}, 1834 (1999).

	\bibitem{Sinova2015}
	J. Sinova, S. O. Valenzuela, J. Wunderlich, C. H. Back, and T. Jungwirth, Spin Hall effects, Rev. Mod. Phys. \textbf{87}, 1213 (2015).

	\bibitem{Saitoh2006}
	E. Saitoh, M. Ueda, H. Miyajima, and G. Tatara, Conversion of spin current into charge current at room temperature: Inverse spin-Hall effect, Appl. Phy. Lett. \textbf{88}, 182509 (2006).

	\bibitem{Ando2011}
	K. Ando \textit{et al}., Inverse spin-Hall effect induced by spin pumping in metallic system, J. Appl. Phys. \textbf{109}, 103913 (2011).

	\bibitem{Manipatruni2019}
	S. Manipatruni, D. E. Nikonov, C.-C. Lin, T. A. Gosavi, H. Liu, B. Prasad, Y.-L. Huang, E. Bonturim, R. Ramesh, and I. A. Young, Scalable energy-efficient magnetoelectric spin-orbit logic, Nature \textbf{565}, 35 (2019).

	\bibitem{Ray1999}
	K. Ray, S. P. Ananthavel, D. H. Waldeck, and R. Naaman, Asymmetric scattering of polarized electrons by organized organic films of chiral molecules, Science \textbf{283}, 814 (1999).

	\bibitem{Gohler2011}
	B. Göhler, V. Hamelbeck, T. Z. Markus, M. Kettner, G. F. Hanne, Z. Vager, R. Naaman, and H. Zacharias, Spin selectivity in electron transmission through self-assembled monolayers of double-stranded DNA, Science \textbf{331}, 894 (2011).

	\bibitem{Nakajima2023}
	R. Nakajima, D. Hirobe, G. Kawaguchi, Y. Nabei, T. Sato, T. Narushima, H. Okamoto, and H. M. Yamamoto, Giant spin polarization and a pair of antiparallel spins in a chiral superconductor, Nature \textbf{613}, 479 (2023).

	\bibitem{Sun2024sciadv}
	R. Sun, Z. Wang, B. P. Bloom, A. H. Comstock, C. Yang \textit{et al.}, Colossal anisotropic absorption of spin currents induced by chirality. Sci. Adv. \textbf{10}, eadn3240 (2024).

	\bibitem{Shiby2025}
	E. Shiby, R. Sun, B. P. Bloom, J. A. Albro, D. Sun and D. H. Waldeck, Interplay between chiro-optical and spin transport properties in chiral CdSe quantum dots. ACS Nano \textbf{19}, 35119 (2025).

	\bibitem{Bloom2024}
	B. P. Bloom, Y. Paltiel, R. Naaman, and D. H. Waldeck, Chiral induced spin selectivity, Chem. Rev. \textbf{124}, 1950 (2024).

	\bibitem{Naaman2019}
	R. Naaman, Y. Paltiel, and D. H. Waldeck, Chiral molecules and the electron spin, Nat. Rev. Chem. \textbf{3}, 250 (2019).

	\bibitem{Xie2011}
	Z. Xie, T. Z. Markus, S. R. Cohen, Z. Vager, R. Gutierrez, and R. Naaman, Spin specific electron conduction through DNA oligomers, Nano Lett. \textbf{11}, 4652 (2011).

	\bibitem{Das2025}
	T. K. Das, N. Preeyanka, S. Mishra, Y. Sang, and C. Fontanesi, Spin-selective anisotropic magnetoresistance driven by chirality in DNA, Adv. Funct. Mater. \textbf{35} 2425377 (2025).

	\bibitem{Mishra2019}
	S. Mishra, S. Pirbadian, A. K. Mondal, M. Y. El-Naggar, and R. Naaman, Spin-dependent electron transport through bacterial cell surface multiheme electron conduits, J. Am. Chem. Soc. \textbf{141}, 19198 (2019).

	\bibitem{Xu2023}
	Y. Xu and W. Mi, Chiral-induced spin selectivity in biomolecules, hybrid organic-inorganic perovskites and inorganic materials: a comprehensive review on recent progress, Mater. Horiz. \textbf{10}, 1924 (2023).

	\bibitem{Qian2022}
	Q. Qian \textit{et al}., Chiral molecular intercalation superlattices, Nature \textbf{606}, 902 (2022).

	\bibitem{Ko2022}
	C.-H. Ko, Q. Zhu, F. Tassinari, G. Bullard, P. Zhang, D. N. Beratan, R. Naaman, and M. J. Therien, Twisted molecular wires polarize spin currents at room temperature, Proc. Natl. Acad. Sci. U.S.A. \textbf{119}, e2116180119 (2022).

	\bibitem{Malatong2023}
	R. Malatong, T. Sato, J. Kumsampao, T. Minato, M. Suda, V. Promarak, and H. M. Yamamoto, Highly durable spin filter switching based on self-assembled chiral molecular motor, Small \textbf{19}, 2302714 (2023).

	\bibitem{Goren2025}
	 N. Goren \textit{et al.}, Antiferromagnetic local volatile memory utilizing non-collinear Mn3Sn thin films and chiral gating. Commun. Mater. \textbf{6}, 238 (2025).

	\bibitem{Al-Bustami2022}
	H. Al-Bustami, S. Khaldi, O. Shoseyov, S. Yochelis, K. Killi, I. Berg, E. Gross, Y. Paltiel, and R. Yerushalmi, Atomic and molecular layer deposition of chiral thin films showing up to 99\% spin selective transport, Nano Lett. \textbf{22}, 5022 (2022).

	\bibitem{Guo2012}
	A.-M. Guo and Q.-F. Sun, Spin-selective transport of electrons in DNA double helix, Phys. Rev. Lett. \textbf{108}, 218102 (2012).

	\bibitem{Guo2014}
	A.-M. Guo and Q.-F. Sun, Spin-dependent electron transport in protein-like single-helical molecules, Proc. Natl. Acad. Sci. U.S.A. \textbf{111}, 11658 (2014).

	\bibitem{Alwan2021}
	S. Alwan and Y. Dubi, Spinterface Origin for the Chirality-Induced Spin-Selectivity Effect, J. Am. Chem. Soc. \textbf{143}, 14235 (2021).

	\bibitem{Das2022}
	T. K. Das, F. Tassinari, R. Naaman, and J. Fransson, Temperature-dependent chiral-induced spin selectivity effect: Experiments and theory, J. Phys. Chem. C \textbf{126}, 3257 (2022).

	\bibitem{Zhang2025a}
	T.-Y. Zhang, Y. Mao, A.-M. Guo, and Q.-F. Sun, Dynamical theory of chiral-induced spin selectivity in electron donor--chiral molecule--acceptor systems, Phys. Rev. B \textbf{111}, 205417 (2025).

	\bibitem{Zhao2025}
	Y. Zhao, K. Zhang, J. Xiao, K. Sun, and B. Yan, Magnetochiral charge pumping due to charge trapping and skin effect in chirality-induced spin selectivity, Nat. Commun. \textbf{16}, 37 (2025).

	\bibitem{Liu2025a}
	P.-Y. Liu, T.-Y. Zhang, and Q.-F. Sun, Dynamical simulation of chiral-induced apin-polarization and magnetization, J. Phys. Chem. Lett. \textbf{16} 6500 (2025).

	\bibitem{Yang2021}
	X. Yang and B. J. Van Wees, Linear-response magnetoresistance effects in chiral systems, Phys. Rev. B \textbf{104}, 155420 (2021).

	\bibitem{Yang2025}
	Q. Yang, Y. Li, C. Felser, and B. Yan, Chirality-induced spin selectivity and current-driven spin and orbital polarization in chiral crystals, Newton \textbf{1}, 100015 (2025).

	\bibitem{Zhang2025b}
	T.-Y. Zhang, Y. Mao, P.-Y. Liu, A.-M. Guo, and Q.-F. Sun, Anomalous magnetoresistance beyond the Jullière model for spin selectivity in chiral molecules, J. Phys. Chem. Lett. \textbf{16} 12596 (2025).

	\bibitem{Gutierrez2012}
	R. Gutierrez, E. Díaz, R. Naaman, and G. Cuniberti, Spin-selective transport through helical molecular systems, Phys. Rev. B \textbf{85}, 081404 (2012).

	\bibitem{Ben2017}
	O. Ben Dor \textit{et al.}, Magnetization switching in ferromagnets by adsorbed chiral molecules without current or external magnetic field, Nat. Commun. \textbf{8}, 14567 (2017).

	\bibitem{Aizawa2023}
	H. Aizawa, T. Sato, S. Maki-Yonekura, K. Yonekura, K. Takaba, T. Hamaguchi, T. Minato, and H. M. Yamamoto, Enantioselectivity of discretized helical supramolecule consisting of achiral cobalt phthalocyanines via chiral-induced spin selectivity effect, Nat. Commun. \textbf{14}, 4530 (2023).

	\bibitem{Adhikari2023}
	Y. Adhikari, T. Liu, H. Wang, Z. Hua, H. Liu, E. Lochner, P. Schlottmann, B. Yan, J. Zhao, and P. Xiong, Interplay of structural chirality, electron spin and topological orbital in chiral molecular spin valves, Nat. Commun. \textbf{14}, 5163 (2023).

	\bibitem{Sun2024}
	R. Sun \textit{et al.}, Inverse chirality-induced spin selectivity effect in chiral assemblies of {$\pi$}-conjugated polymers, Nat. Mater. \textbf{23}, 782 (2024).

	\bibitem{Ando2013}
	K. Ando, S. Watanabe, S. Mooser, E. Saitoh, and H. Sirringhaus, Solution-processed organic spin-charge converter, Nat. Mater. \textbf{12}, 622 (2013).

	\bibitem{Qiu2015}
	Z. Qiu, M. Uruichi, D. Hou, K. Uchida, H. M. Yamamoto, and E. Saitoh, Spin-current injection and detection in {$\kappa$}-(BEDT-TTF)2Cu[N(CN)2]Br, AIP Advances \textbf{5}, 057167 (2015).

	\bibitem{Inui2020}
	A. Inui \textit{et al.}, Chirality-induced spin-polarized state of a chiral crystal CrNb3S6, Phys. Rev. Lett. \textbf{124}, 166602 (2020).

	\bibitem{Shiota2021}
	K. Shiota \textit{et al.}, Chirality-induced spin polarization over macroscopic distances in chiral disilicide crystals, Phys. Rev. Lett. \textbf{127}, 126602 (2021).

	\bibitem{Park2022}
	K. S. Park, Z. Xue, B. B. Patel, H. An, J. J. Kwok, P. Kafle, Q. Chen, D. Shukla, and Y. Diao, Chiral emergence in multistep hierarchical assembly of achiral conjugated polymers, Nat. Commun. \textbf{13}, 2738 (2022).

	\bibitem{Nabei2020}
	Y. Nabei, D. Hirobe, Y. Shimamoto, K. Shiota, A. Inui, Y. Kousaka, Y. Togawa, and H. M. Yamamoto, Current-induced bulk magnetization of a chiral crystal CrNb3S6, Appl. Phy. Lett. \textbf{117}, 052408 (2020).

	\bibitem{SM}
	See Supplemental Material at \url{https://doi.org/10.1103/xskm-5w1g} for more details, which includes Refs. \cite{Al-Bustami2022,Sun2005b,Chen2024,Zhang2025b,Datta1995,Xing2007,Sun2024,Dong2025}.

	\bibitem{Song2007}
	H. Song, H. Lee, and T. Lee, Intermolecular chain-to-chain tunneling in metal-alkanethiol-metal junctions, J. Am. Chem. Soc. \textbf{129}, 3806 (2007).

	\bibitem{Frederiksen2009}
	T. Frederiksen, C. Munuera, C. Ocal, M. Brandbyge, M. Paulsson, D. Sanchez-Portal, and A. Arnau, Exploring the tilt-angle dependence of electron tunneling across molecular junctions of self-assembled alkanethiols, ACS Nano \textbf{3}, 2073 (2009).

	\bibitem{Liu2025b}
	P.-Y. Liu, T.-Y. Zhang, A.-M. Guo, Y. Paltiel, and Q.-F. Sun, Spin-to-charge conversion modulated by chiral molecules, J. Phys. Chem. Lett. \textbf{16} 10426 (2025).

	\bibitem{Moharana2025}
	A. Moharana, Y. Kapon, F. Kammerbauer, D. Anthofer, S. Yochelis, H. Shema, E. Gross, M. Kläui, Y. Paltiel, and A. Wittmann, Chiral-induced unidirectional spin-to-charge conversion, Sci. Adv. \textbf{11}, eado4285 (2025).

	\bibitem{Brataas2002}
	A. Brataas, Y. Tserkovnyak, G. E. W. Bauer, and B. I. Halperin, Spin battery operated by ferromagnetic resonance, Phys. Rev. B \textbf{66}, 060404 (2002).

	\bibitem{Wang2004}
	D.-K. Wang, Q.-F. Sun, and H. Guo, Spin-battery and spin-current transport through a quantum dot, Phys. Rev. B \textbf{69}, 205312 (2004).

	\bibitem{Sun2005a}
	Q.-F. Sun and X. C. Xie, Definition of the spin current: The angular spin current and its physical consequences, Phys. Rev. B \textbf{72}, 245305 (2005).

	\bibitem{Edelstein1990}
	V. M. Edelstein, Spin polarization of conduction electrons induced by electric current in two-dimensional asymmetric electron systems, Solid State Commun. \textbf{73}, 233 (1990).

	\bibitem{Bychkov1984}
	Y. A. Bychkov and E. I. Rashba, Oscillatory effects and the magnetic susceptibility of carriers in inversion layers, J. Phys. C: Solid State Phys. \textbf{17}, 6039 (1984).

	\bibitem{Pan2016}
	T.-R. Pan, A.-M. Guo, and Q.-F. Sun, Spin-polarized electron transport through helicene molecular junctions, Phys. Rev. B \textbf{94}, 235448 (2016).

	\bibitem{Wang2024}
	C. Wang, Z.-R. Liang, X.-F. Chen, A.-M. Guo, G. Ji, Q.-F. Sun, and Y. Yan, Transverse spin selectivity in helical nanofibers prepared without any chiral molecule, Phys. Rev. Lett. \textbf{133}, 108001 (2024).

	\bibitem{Steele2013}
	G. A. Steele, F. Pei, E. A. Laird, J. M. Jol, H. B. Meerwaldt, and L. P. Kouwenhoven, Large spin-orbit coupling in carbon nanotubes, Nat. Commun. \textbf{4}, 1573 (2013).

	\bibitem{Sun2005b}
	Q.-F. Sun, J. Wang, and H. Guo, Quantum transport theory for nanostructures with Rashba spin-orbital interaction, Phys. Rev. B \textbf{71}, 165310 (2005).

	\bibitem{Chen2024}
	S. Chen, R. Wu, and H.-H. Fu, Persistent chirality-induced spin-selectivity effect in circular helix molecules, Nano Lett. \textbf{24}, 6210 (2024).

	\bibitem{Xing2007}
	Y. Xing, Q.-F. Sun, and J. Wang, Symmetry and transport property of spin current induced spin-Hall effect, Phys. Rev. B \textbf{75}, 075324 (2007).

	\bibitem{Lee1981}
	D. H. Lee and J. D. Joannopoulos, Simple scheme for surface-band calculations. I, Phys. Rev. B \textbf{23}, 4988 (1981).

	\bibitem{Datta1995}
	S. Datta, \textit{Electronic Transport in Mesoscopic Systems}, 1st ed. (Cambridge University Press, 1995).

	\bibitem{Yan2002}
	Y. J. Yan and H. Zhang, Toward the mechanism of long-range charge transfer in DNA: theories and models, J. Theor. Comput. Chem. \textbf{01}, 225 (2002).

	\bibitem{Dong2025}
	Y. Dong, A. McConnell, M. P. Hautzinger, M. A. Haque, A. H. Comstock, P. M. Theiler, J. M. Luther, P. C. Sercel, D. Sun, and M. C. Beard, Ultrafast inverse chirality-induced spin selectivity observed by THz emission, Science \textbf{390}, 595 (2025).

	\bibitem{Mishra2020}
	S. Mishra, A. K. Mondal, S. Pal, T. K. Das, E. Z. B. Smolinsky, G. Siligardi, and R. Naaman, Length-dependent electron spin polarization in oligopeptides and {DNA}, J. Phys. Chem. C \textbf{124}, 10776 (2020).

	\bibitem{Evers2022}
	F. Evers \textit{et al.}, Theory of chirality induced spin selectivity: Progress and challenges, Advanced Materials \textbf{34}, 2106629 (2022).

	\bibitem{Jhang2010}
	S. H. Jhang, M. Marganska, Y. Skourski, D. Preusche, B. Witkamp, M. Grifoni, H. Van Der Zant, J. Wosnitza, and C. Strunk, Spin-orbit interaction in chiral carbon nanotubes probed in pulsed magnetic fields, Phys. Rev. B \textbf{82}, 041404 (2010).



\end{thebibliography}
\end{document}